\documentclass[aps,pra,pdflatex,reprint,10pt,superscriptaddress,floatfix,nofootinbib]{revtex4-2}

\usepackage{float}
\usepackage{amsmath}
\usepackage[T1]{fontenc}
\usepackage[utf8]{inputenc}
\usepackage{lmodern}
\usepackage{amssymb}
\usepackage{natbib}
\usepackage{graphicx}
\usepackage{hyperref}
\usepackage{epstopdf}
\usepackage{pifont}
\usepackage{textgreek}
\usepackage[normalem]{ulem}
\usepackage{placeins}
\usepackage{tikz}
\usepackage{physics}
\usepackage{orcidlink}
\usepackage{bm}

\graphicspath{{figures/}}

\hypersetup{
    colorlinks=true,
    linkcolor=blue,
    citecolor=blue,
    urlcolor=blue
}

\begin{document}

\title{Torsional selection rule for the spin--orbit conversion of light}

\author{Edilberto O. Silva\orcidlink{0000-0002-0297-5747}}
\email[Edilberto O. Silva - ]{edilberto.silva@ufma.br}
\affiliation{
Programa de P\'os-Gradua\c c\~ao em F\'{\i}sica \&
Coordena\c c\~ao do Curso de F\'{\i}sica -- Bacharelado,
Universidade Federal do Maranh\~{a}o,
65085-580 S\~{a}o Lu\'{\i}s, Maranh\~{a}o, Brazil}

\date{\today}

\begin{abstract}
Standard Pancharatnam--Berry and linear-birefringent media convert optical
spin into orbital angular momentum (OAM) through an anisotropy
\emph{director}, a rank-two, headless field, and therefore obey the
selection rule $\Delta\ell=2q$ per unit texture charge $q$. We show that a
medium with geometric \emph{torsion}, the continuum limit of a
screw-dislocation array, can convert spin to OAM through the
\emph{contortion} of its material connection, which enters the effective
paraxial dynamics as a rank-one vector field. The resulting selection rule is
$\Delta\ell=q$. Its winding is fixed by geometry and symmetry, not by a
Pancharatnam--Berry director, and the process conserves the screw charge
$\tilde J_z=L_z+(q/2)\sigma_z$ while exchanging $(2-q)\hbar$ of angular
momentum per converted photon with the defect lattice. Paraxial simulations
confirm the rule: a circular Gaussian input develops a stable, topologically
quantized $\ell=+q$ vortex in the reversed helicity, with $83\%$ conversion
over three Rayleigh ranges and no fine-tuning. We propose a polarization-resolved photonic-lattice discriminator in which the slope of the measured OAM
versus the independently written texture charge, one for torsion, two for
birefringence, separates the two mechanisms.
\end{abstract}

\maketitle

\emph{Introduction. }The interconversion of the spin and orbital angular
momentum (OAM) of light~\cite{Allen1992,LibermanZeldovich1992,Bliokh2008,
Bliokh2015,Shen2019,Forbes2021} is the working principle of a wide class of
structured-light devices. More broadly, the OAM spectrum provides an
infinite-dimensional photonic degree of freedom that can be engineered for
multidimensional angular-momentum states and used as a spiral spectrum to
probe phase, amplitude, and dislocation features of matter~\cite{MolinaTerriza2002,Torner2005}.
Its paradigm is the $q$-plate: an anisotropic
element whose optical axis winds $q$ times around a central defect and which
flips the helicity of a photon while imprinting an OAM jump
$\Delta\ell=2q$~\cite{Pancharatnam1956,Berry1984,Marrucci2006,Marrucci2011,
Rubano2019}, a rule that carries over to dielectric
metasurfaces~\cite{Devlin2017} and, more generally, to any medium whose
spin--orbit coupling is mediated by linear birefringence. The factor of two is
not accidental: the local optic axis is a \emph{director}, a headless,
rank-two object, so its optical response winds twice per turn of the
texture. This rank-two rule underlies the standard Pancharatnam--Berry route
to spin-to-OAM conversion.

\begin{figure}[t]\centering
\includegraphics[width=\linewidth]{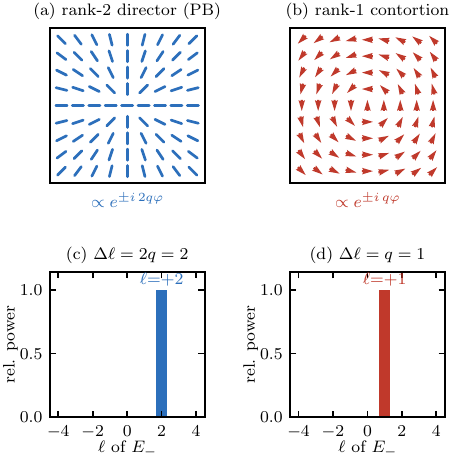}
\caption{The selection rule distinguishes the rank of the mediating field.
(a) A Pancharatnam--Berry element couples the circular components through a
rank-two director texture of charge $q$; the coupling winds as
$e^{\pm i2q\varphi}$. (b) A torsional (Riemann--Cartan) medium couples them
through the rank-one contortion vector, which winds as $e^{\pm iq\varphi}$.
(c),(d) Simulated OAM spectra of the converted beam $E_-$ for the same
texture charge $q=1$ and identical coupling magnitude: the sideband appears
at $\ell=+2$ (PB) versus $\ell=+1$ (torsional).}
\label{fig:mechanism}
\end{figure}

Here we identify a physically distinct class of converters that does not. In
the geometric theory of defects~\cite{Bilby1955,Kroner1962,Katanaev1992,
Puntigam1997}, a continuous distribution of parallel screw dislocations is
described by a Riemann--Cartan geometry with \emph{torsion}: curvature-free,
but with a nontrivial translational holonomy. Light in such media acquires
quantized OAM modes~\cite{Fumeron2015}, dislocation-induced anisotropy that
alters its spin~\cite{Mashhadi2010}, a purely geometric optical activity
linear in the dislocation density~\cite{BelichSilva2026,Zhang2014}, and
torsion-controlled guidance~\cite{GurtasDogan2025}; torsion-stress-induced
vortices have been observed in bulk crystals~\cite{Skab2011,Vasylkiv2013},
and photonic lattices emulating dislocated backgrounds bind vortex
light~\cite{Sheng2022}. We show that the object mediating spin--orbit
conversion in a torsional medium is the transverse part of the
\emph{contortion} tensor, a rank-one (vector) field, that winds
\emph{once} per turn of the defect texture. The selection rule is therefore
\begin{equation}
  \Delta\ell_{\mathrm{torsional}} = q,
  \qquad
  \Delta\ell_{\mathrm{PB}} = 2q,
  \label{eq:rule}
\end{equation}
per converted photon (Fig.~\ref{fig:mechanism}). The slope of the measured
OAM versus texture charge is thus a clean, binary discriminator between
geometric (torsional) and birefringent conversion, a testable prediction
that we quantify below for existing photonic platforms.

\begin{figure*}[t]\centering
\includegraphics[scale=0.98]{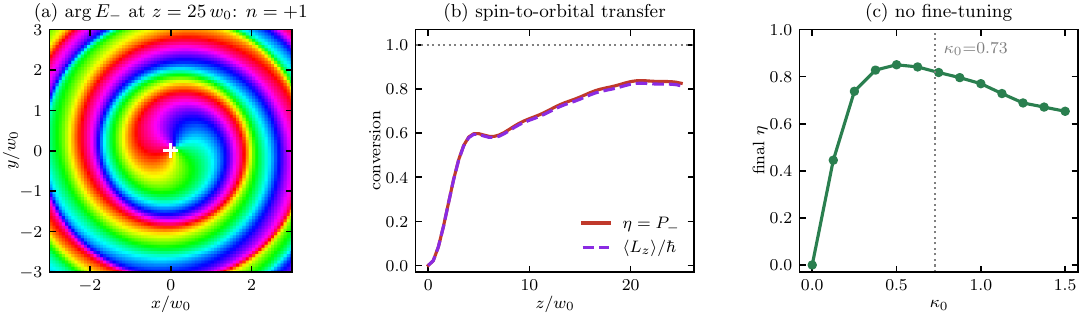}
\caption{Torsional vortex generation ($q=1$). (a) Phase of the converted
field at $z=25\,w_0$: a stable, charge-$n{=}{+}1$ spiral. (b) Conversion
$\eta=P_-$ and mean OAM versus propagation distance; both saturate at
$0.83$. (c) Final conversion versus holonomy strength $\kappa_0$: a broad
maximum ($\eta\simeq0.85$) with no fine-tuning; in every run the charge is
quantized at $n=+1$. Full sweeps in $(\kappa_0,\Gamma_0)$, paraxial parameter
$k_0=8$--$64$, and grid refinement are given in the SM~\cite{SuppMat}.}
\label{fig:results}
\end{figure*}

\emph{Model. }We describe a monochromatic paraxial beam by the
circular-polarization spinor $\Psi=(E_+,E_-)^{\mathsf T}$ propagating along
the defect axis $z$ of a screw-dislocation continuum,
\begin{equation}
  i\,\partial_z \Psi
  = \Bigl[ -\tfrac{1}{2k_0}\nabla_\perp^2
    + \Gamma(r)\,\sigma_z
    + \kappa(r)\,\hat{\bm w}\!\cdot\!\bm\sigma_\perp \Bigr]\Psi ,
  \label{eq:model}
\end{equation}
with $\hat{\bm w}\!\cdot\!\bm\sigma_\perp
=\cos(q\varphi)\sigma_x+\sin(q\varphi)\sigma_y$, $k_0=kw_0$, and transverse
lengths in units of the input waist $w_0$. Equation~\eqref{eq:model} is an effective paraxial Maxwell model for a
structured medium; it is not an assumption of a universal vacuum coupling of
photons to spacetime torsion. The material response projects two independent
components of the contortion one-form of the dislocated medium, computed in
the Supplemental Material (SM)~\cite{SuppMat}, onto the circular-polarization
subspace: the component along $e^{\hat z}$ rotates the transverse
frame and gives the diagonal torsional phase rate $\Gamma(r)\propto A'(r)$,
reproducing the geometric optical activity law of the uniform-torsion
medium~\cite{BelichSilva2026}; the transverse components form the contortion vector
$\bm w=\tfrac{\tau}{2}\hat{\bm\varphi}$, where $\tau$ is the torsion
density. The tangential direction differs from the convention used in
Eq.~\eqref{eq:model} only by a constant $\pi/2$ phase, which can be absorbed
into the circular-spin basis and does not affect the winding or the selection
rule. The azimuthal winding is exactly $q=1$ for the axisymmetric screw
continuum and $q$ for engineered Burgers textures of integer charge $q$.
Because $\bm w$ is a vector, the symmetry-allowed local Hermitian coupling
that is first order in the contortion carries the phase $e^{\pm iq\varphi}$
in Eq.~\eqref{eq:model}; rank-two combinations
$w_iw_j-\tfrac12\delta_{ij}\bm w^2$ wind as $e^{\pm i2q\varphi}$ and
reproduce the director-mediated PB coupling instead. The holonomy amplitude
$\kappa(r)$ vanishes on axis, as required by regularity, and saturates over a
scale $R_H$; its magnitude is not universal, but is set by the torsion density
and by the magnetoelectric/elasto-optic response of the host~\cite{Zhang2014,
Mashhadi2010} (SM~\cite{SuppMat}).

\begin{figure*}[t]\centering
\includegraphics[scale=0.98]{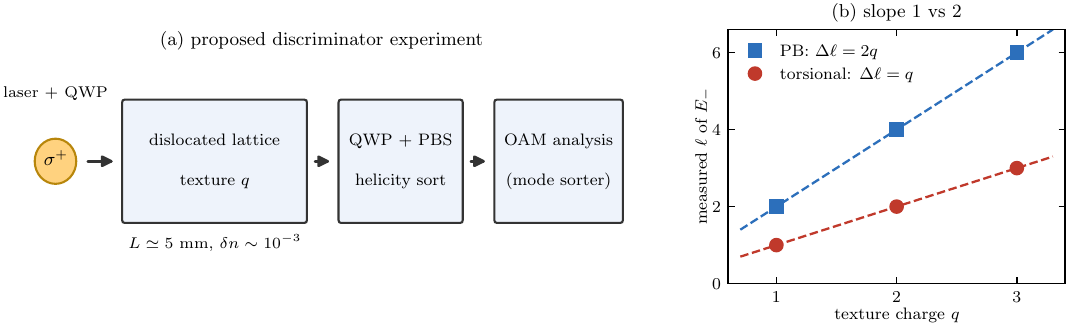}
\caption{Proposed discriminator. (a) A circularly polarized Gaussian
traverses a femtosecond-written lattice with a screw-dislocation texture of
charge $q$ ($L\simeq5\,$mm at $\delta n\sim10^{-3}$); the output is
helicity-sorted and its OAM measured. (b) Predicted OAM of the converted
beam versus texture charge, from full simulations of Eq.~\eqref{eq:model}
(circles) and of a PB-type rank-two coupling of identical magnitude
(squares): slopes $1$ and $2$, with no overlap at any integer $q$.}
\label{fig:experiment}
\end{figure*}
The off-diagonal term feeds $E_-$ from $E_+$ with the factor $e^{+iq\varphi}$:
an input $E_+\propto e^{im\varphi}$ generates $E_-\propto e^{i(m+q)\varphi}$,
which is Eq.~\eqref{eq:rule}. The rule is protected by symmetry. The
generator of Eq.~\eqref{eq:model} commutes with the \emph{screw charge}
\begin{equation}
  \tilde J_z = L_z + \tfrac{q}{2}\,\sigma_z, \qquad L_z=-i\partial_\varphi,
  \label{eq:charge}
\end{equation}
the exact conserved quantity of the torsional medium; input
($\sigma_z{=}{+}1$, $\ell{=}0$) and converted ($\sigma_z{=}{-}1$, $\ell{=}q$)
states both carry $\tilde J_z=q/2$. The \emph{optical} angular momentum is
not conserved: each converted photon loses $2\hbar$ of spin and gains only
$q\hbar$ of OAM, the balance $(2-q)\hbar$ being exchanged as a mechanical torque with the
dislocation lattice, the torsional analogue of the angular-momentum exchange
in a $q$-plate, for which only $q=1$ is conserving~\cite{Marrucci2006,
Marrucci2011}. Here the conserving texture is instead $q=2$, a second,
independent signature of the rank-one mechanism.

\emph{Vortex generation. }We integrate Eq.~\eqref{eq:model} with a
symmetric split-step scheme (parameters and convergence tests in
SM~\cite{SuppMat}; baseline $k_0=8$, $\Gamma_0=0.8$, $\kappa_0=0.73$, $q=1$,
propagation to $z=25\,w_0\approx3z_R$). A circularly polarized Gaussian
($\ell=0$) develops, in the reversed helicity, a phase singularity that
nucleates on axis within a fraction of a Rayleigh range and locks to
topological charge $n=+1$ [Fig.~\ref{fig:results}(a)]. The conversion is
quantitative: the power fraction in the reversed helicity and the mean OAM
grow together and saturate at $\eta=P_-\simeq0.83$,
$\langle L_z\rangle\simeq0.83\,\hbar$ [Fig.~\ref{fig:results}(b)]. Here
$\langle L_z\rangle$ is normalized to the total input power; within the
converted channel the mean OAM per photon is $\ell=+1$. The identity
$\langle L_z\rangle/\hbar=\eta$ is therefore the macroscopic expression of
the selection rule, and the OAM spectrum of $E_-$ is a single line at
$\ell=+1$ with purity above $0.999$. Switching off the holonomy ($\kappa\equiv0$) leaves
the full Abelian torsional phase $\Gamma\sigma_z$ in place, yet produces
\emph{no} conversion and no vortex (SM~\cite{SuppMat}): the rank-one
contortion coupling is the sole microscopic agent.

\emph{Robustness. }The quantization and chirality of the output are
structural, not tuned. Across more than one hundred simulations a
13-point sweep of $\kappa_0$ [Fig.~\ref{fig:results}(c)], a $9\times9$
landscape in $(\kappa_0,\Gamma_0)$, and paraxial parameters up to $k_0=64$
($w_0\approx10\lambda$) the topological charge is $n=+1$ and the recorded
azimuthal purity exceeds $0.999$ in every run, while only the efficiency
varies smoothly ($\eta$ up to $0.86$; $\Delta\eta<0.2\%$ under doubling of
grid or steps)~\cite{SuppMat}. The rule $\Delta\ell=q$ cannot be washed out
by parameter drift: the off-diagonal coupling can only populate the
$\ell=m+q$ channel.

\emph{Proposed experiment. }Figure~\ref{fig:experiment} shows a
discriminator experiment built from demonstrated components.
Femtosecond-written photonic lattices already emulate dislocated
backgrounds and bind vortex light~\cite{Sheng2022}. The required extra
ingredient is a polarization-sensitive implementation in which the local
waveguide anisotropy or bi-anisotropy is aligned with the transverse Burgers
texture, so that the circular--polarization coupling is linear in the written
contortion vector of Eq.~\eqref{eq:model}; this can be implemented, for
example, with elliptic femtosecond-written waveguides, controlled stress
birefringence, or locally anisotropic couplers. Such laser-written platforms
provide effective index contrasts $\delta n\sim10^{-3}$. The accumulated rotation required for the $83\%$
operating point is $\kappa_{\rm dim}L\approx18$, i.e.,
$L\approx4.6\,$mm at $\lambda=800\,$nm for $\delta n=10^{-3}$, a
centimeter-class chip; twisted photonic-crystal
fibers~\cite{Russell2017} and liquid-crystal realizations of dislocated
media~\cite{Fumeron2015} give $\delta n\sim10^{-4}$ and centimeter lengths.
A $\sigma^+$ Gaussian is launched along the defect axis; the output is
helicity-sorted (quarter-wave plate and polarizing splitter) and the OAM
content of the reversed-helicity channel is read out by a mode sorter or
tilted-lens diagnostics. The smoking gun is the slope of the measured
$\ell$ versus the independently written texture charge $q$
[Fig.~\ref{fig:experiment}(b)]: unity for the torsional mechanism, two for a
birefringence-mediated (PB) response. Although PB plates may employ half-integer director charges, the proposed test compares devices with the same
integer vector texture $q$, for which the $\ell=q$ and $\ell=2q$ sidebands do
not overlap. Reversing either the input helicity or the sign of the Burgers
texture must reverse the sideband, $\ell\to-\ell$, providing an additional
control against residual PB leakage. Two corroborating signatures come for free: for $q=2$ the torsional
conversion is angular-momentum conserving, so no torque is exerted on the
lattice, whereas the PB response at $q=2$ requires an equal-and-opposite
exchange of $2\hbar$ per converted photon with the medium; and the converted
power follows the distributed (Rabi-like) growth of
Fig.~\ref{fig:results}(b) rather than the thin-element retardation law of a
$q$-plate.

\emph{Discussion. }Equation~\eqref{eq:rule} elevates the OAM jump per unit
texture charge to a measurement of the \emph{rank} of the geometric object
that couples light's spin to its spatial structure: directors give two,
vectors give one. Torsion provides the natural physical realization of the
rank-one case: the contortion of a Riemann--Cartan medium is a transverse
vector field whose winding is fixed by the defect topology, not by a material
director. The associated conserved charge~\eqref{eq:charge}, the mechanical
torque $(2-q)\hbar$ per converted photon, and the distributed conversion
dynamics complete a phenomenology that is qualitatively distinct from the
Pancharatnam--Berry paradigm. Beyond structured light, the effect gives
photonics a direct probe of material torsion: the conversion efficiency and sideband
position measure the dislocation density and texture charge of the medium,
complementing recent proposals for torsion metrology in electronic
systems~\cite{Silva2026,Baimuratov2015} and extending the toolbox of
geometric and topological photonics~\cite{Pendry2006,Leonhardt2006,
Sheng2013,Bekenstein2015,Ozawa2019} to non-Riemannian backgrounds.

\begin{acknowledgments}
The author thanks Lluís Torner for stimulating correspondence on the
inverse characterization of torsional optical media from spiral spectra.
This work was supported by CAPES (Finance Code 001), CNPq (Grant
306308/2022-3), and FAPEMA (Grants UNIVERSAL-06395/22 and APP-12256/22).
\end{acknowledgments}

\end{document}